\documentclass[12pt]{article}
\usepackage[utf8]{inputenc}
\usepackage[a4paper, margin=1in]{geometry}
\usepackage{graphicx}
\usepackage{amsmath, amssymb}
\usepackage{mhchem}
\usepackage{setspace}
\usepackage{xcolor}
\usepackage{cite}
\usepackage{hyperref}
\usepackage{titlesec}
\usepackage{float}
\usepackage{tikz}
\usepackage{longtable}  
\usetikzlibrary{positioning, arrows.meta}
\usepackage{appendix}
\titleformat{\section}{\large\bfseries}{\thesection.}{1em}{}
\titleformat{\subsection}{\normalsize\bfseries}{\thesubsection.}{1em}{}

\title{\textbf{A generic network theoretic based model to classify SDG indicators}}
\author{
  Gaurav Kottari\thanks{Email: gk917@snu.edu.in} \\
  \small Department of Mathematics \\
  \small Shiv Nadar Institution of Eminence\\
  \small Delhi-NCR, India
  \and
  Qazi J. Azhad\thanks{Corresponding Email: qazi.jamal@snu.edu.in, qaziazhadjamal@gmail.com} \\
  \small Department of Mathematics\\
  \small Shiv Nadar Institution of Eminence\\
  \small Delhi-NCR, India
  \and
  Niteesh Sahni\thanks{Email: niteesh.sahni@snu.edu.in} \\
  \small Department of Mathematics\\
  \small Shiv Nadar Institution of Eminence\\
  \small Delhi-NCR, India
}
\date{\today}

\begin{document}

\maketitle

\begin{abstract}
To achieve the United Nations Sustainable Development Goals, coordinated action across their interlinked indicators is required. Although most of the research on the interlinkages of the SDGs is done at the goal level, policies are usually made and implemented at the level of indicators (or targets). Our study examines the existing literature on SDG interlinkages and indicator (or target) prioritization, highlighting important drawbacks of current methodologies. To address these limitations, we propose a generic network-based model that can quantify the importance of the SDG indicators and help policymakers in identifying indicators for maximum synergistic impact. Our model applies to any country, offering a tool for national policymakers. We illustrate the application of this model using data from India, identifying important indicators that are crucial for accelerating progress in the SDGs. While our main contribution lies in developing this network-theoretic methodology, we also provide supporting empirical evidence from existing literature for selected key observations. 
\end{abstract}


\section{Introduction}
Sustainable Development Goals (SDGs) were adopted by the member states of the United Nations in 2015, which is a universal call to eradicate poverty, protect the environment, and guarantee peace and prosperity for all by 2030~\cite{assembly2015resolution}. It addresses a wide range of social, economic, and environmental issues and consists of 17 goals and 169 related targets. Progress toward one goal rarely occurs in isolation because these objectives are wide-ranging and interrelated. Improvements in fields such as health, education, or clean energy can support successes in other fields, but they can also lead to unexpected setbacks or new pressures in other areas. Therefore, identifying these interdependencies is essential for understanding the overall dynamics of sustainable development.

Since each goal is further broken down into specific targets that guide concrete actions and measurement, it becomes important to study interactions not only among goals but also among their underlying targets, where the most direct trade-offs and synergies often emerge. Analyzing the interlinkages between the targets is essential to find synergies that can speed up progress, and trade-offs that could impede it~\cite{allen2019prioritising}. 

To address this need for understanding interlinkages, existing studies have employed two broad approaches: qualitative and quantitative. In qualitative studies, the strength of the interlinkages between targets is determined by expert opinions, policy analysis, or literature reviews~\cite{allen2019prioritising,nilsson2016policy,weitz2018towards,tremblay2020sustainable,pham2020interactions,xiao2023synergies,fader2018toward}. In contrast, quantitative studies establish these strengths of interlinkage in a data-driven manner, often using techniques such as pairwise correlation analysis or principal component analysis on indicator-level datasets~\cite{pradhan2017systematic,de2020synergies,hegre2020synergies,kostetckaia2022sustainable,miao2025priority}.

Because policy interventions are made at the level of specific indicators, determining which of these are synergistic becomes essential for maximizing development gains. By “synergistic”, we mean indicators whose progress tends to produce net positive effects across other indicators, where the positive spillovers from improving these indicators outweigh any negative effects on others. Since interventions on these indicators reinforce progress not only within their immediate domain but also across multiple interconnected areas, identifying such indicators guarantees that available resources are invested in generating the maximum systemic payoffs. Also, finding indicators that entail substantial trade-offs is equally important. If taken carelessly, actions based on these indicators could unintentionally impede or even reverse progress in other areas. Policymakers can create strategies that maximize the overall efficiency and effectiveness of sustainable development efforts by strengthening complementarities while minimizing potential conflicts because they are aware of the indicators that generate trade-offs as well as those that reinforce synergies. 

Ranganathan and Swain~\cite{swain2021modeling} proposed a methodology using network theory to prioritize SDG indicators. They constructed regional correlational networks of SDG indicators and applied a medium correlation threshold of $0.5$, thus considering both strong synergies (correlations $> 0.5$) and strong trade-offs (correlations $< –0.5$). Their results showed that strong negative correlations were absent in the four global regions (OECD, East Asia, Latin America, and MENA). Only one was observed in Sub-Saharan Africa, and two in South Asia. Since the net impact of trade-offs appeared to be weak, they used centrality measures to identify the most influential indicators in each regional network. However, when we applied this methodological approach to the data set for India, we observed a substantial number of trade-offs, many with significant negative correlations (below $–0.9$). This contradicts the regional results revealed in their study~\cite{swain2021modeling}, where such strong trade-offs were rarely observed. This observation suggests that while the method proposed by Ranganathan and Swain~\cite{swain2021modeling} is useful in broader regional contexts, it may not be directly applicable to national or local-level decision-making.

In 2023, Song and Jang~\cite{song2023unpacking} suggested another framework to rank SDG targets, by constructing a network based on the similarity between target keywords using semantic analysis. Their approach only establishes a link when two targets share moderate semantic similarities. This method, however, overlooks the possibility of trade-offs between targets, which are crucial for understanding the full range of interactions.

The drawbacks observed at the national level highlight the need for a more flexible method of identifying important indicators. In this paper, we propose a generic model that addresses key limitations of existing approaches. Our methodology allows more flexibility and adaptability across several nations by not depending on fixed correlation cutoffs. In addition, it explicitly captures both synergies and trade-offs between indicators, allowing a more comprehensive and country-specific classification of SDG indicators into synergy-dominated and trade-off-dominated.

In this paper, we make several contributions. First, we formalize the definition of when an indicator can be considered synergy-dominated or trade-off–dominated. To the best of our knowledge, such a measure has not been proposed in the existing literature, where researchers have mainly focused on quantifying the pairwise strength of relationships between indicators. Our formulation instead provides a way to assess the overall systemic influence of a particular indicator across the network.

Second, to demonstrate the practical utility of the model, we conduct a case study using SDG indicator data for India. We classify the indicators into synergy-dominated and trade-off–dominated categories, highlighting those that generate strong positive spillovers and those that create widespread negative effects. Notably, we find that access to piped water is trade-off–dominated: investments in this indicator exhibit negative correlations with several other indicators, especially those related to health. This suggests that while improving piped water coverage is valuable in itself, it may simultaneously introduce new systemic challenges. Finally, we validate our findings by comparing them with existing sector-specific research. In particular, studies focusing on piped water provision report similar adverse interactions with health outcomes, supporting the robustness of our model’s results.

The structure of this paper is as follows. Section~\ref{3 section2} outlines the construction of complete weighted SDG networks, introduces measures of synergy and trade‑off strengths, and explains the logistic regression framework using direct and indirect network effects to classify and rank targets. Section~\ref{3 section 3} applies this framework by building a logistic regression model and demonstrating its application through a detailed case study on India. Section~\ref{3 section 4} concludes the study by summarizing the findings and discussing its limitations.

\section{Data and Methodology}\label{3 section2}

\subsection{Data}
In our study, we used the dataset from the Sustainable Development Report 2025, which we obtained from the official SDG Index website (\url{https://dashboards.sdgindex.org/downloads}). The dataset includes scores for SDG indicators related to the 17 Sustainable Development Goals across different countries, covering the period from 2000 to 2024. Each indicator value ranges from 0 to 100. A score of 100 means the indicator has been achieved, while a score of 0 indicates the poorest performance observed among all countries. These scores enable a comparison of the SDG indicators.

\subsection{Methodology}

\subsubsection{Network Theory and Measures}\label{3 sec:network-measures}

Here, we recall the basic concepts of network theory used in this study and introduce the measures used to characterize node-level interactions based on edge weights.

To provide a foundation for our analysis, we first recall standard definitions from network theory. These definitions are well established and can be found in~\cite{estrada2015first,newman2018networks}. A network (or graph) $G = (V, E)$ is an ordered pair consisting of a finite set $V$ of vertices (or nodes) and a set $E$ of edges, where $E \subseteq [V]^2$. Here, $[V]^2$ denotes the set of all two-element subsets of $V$. If $E = [V]^2$, then $G$ is called a complete network. Each edge $e_{ij} = \{i, j\}$ of the network $G$ may be associated with a real number $w_{ij}$, referred to as its weight. The network $G$, together with these edge weights $W = \{\, w_{ij} \mid i,j \in V \,\}$, is called a weighted network. If no weights are assigned to the edges, the network is referred to as an unweighted network. In such cases, each edge is conventionally assumed to have a weight of one.

\begin{figure}[h]
    \centering
    \includegraphics[width=0.95\textwidth]{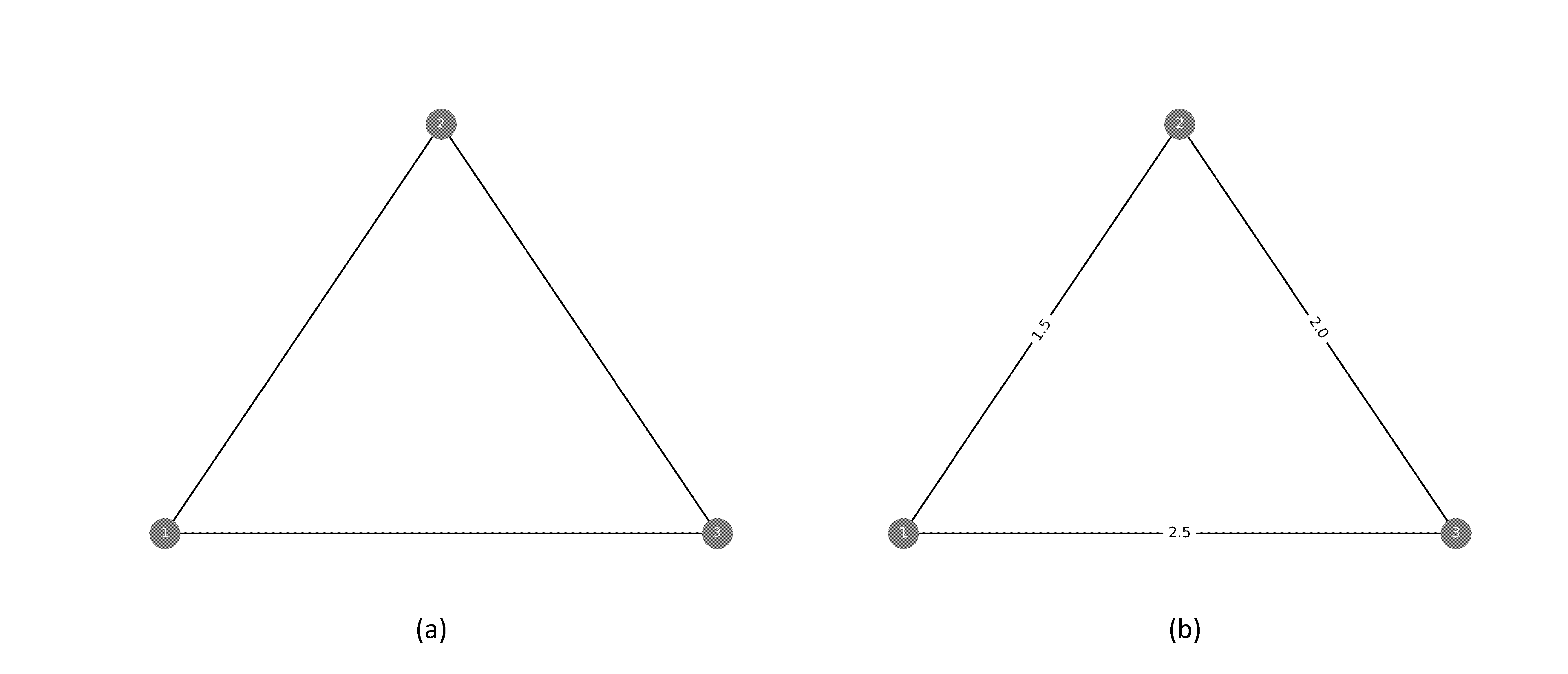}
    \caption{(a) An unweighted complete network with three vertices. (b) A weighted complete network on the same set of vertices, where each edge is assigned a real-valued weight.}
    \label{fig:example-graph}
\end{figure}

To illustrate these definitions, Figure~\ref{fig:example-graph} presents an example of a complete unweighted graph and a complete weighted graph with three vertices.

Let $G = (V, E, W)$ denote a weighted network, where the edge weights $w_{ij}$ can take both positive and negative values, representing, for instance, supportive or opposing relationships between nodes.

For any node $i \in V$, we define a star subgraph, denoted by $G_i$. The star subgraph $G_i$ consists of the node $i$ itself and all edges that are directly connected to it. In other words, it includes $i$ and all the nodes that share an edge with $i$.

We further divide this star subgraph into two parts:
\begin{itemize}
    \item $G_i^+$: the subgraph containing the vertex $i$ and only the positive weighted edges incident on it, i.e., edges where $w_{ij} > 0$; and
    \item $G_i^-$: the subgraph containing the vertex $i$ and only the negative weighted edges incident on it, i.e., edges where $w_{ij} < 0$.
\end{itemize}

\begin{figure}[h!]
    \centering
    \includegraphics[width=0.95\textwidth]{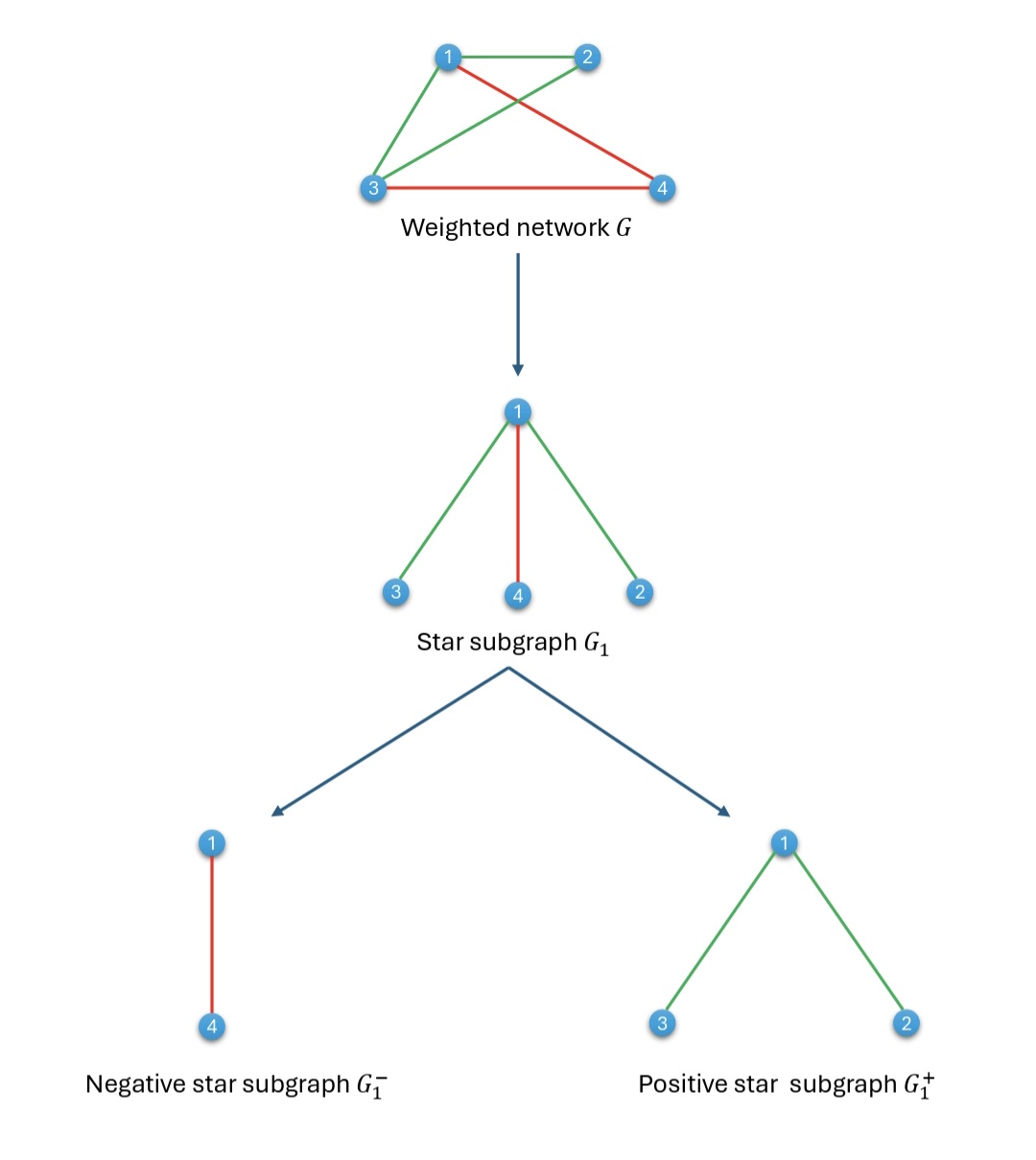}
    \caption{Illustration of a weighted network $G$ with four nodes, its star subgraph $G_{1}$ corresponding to node $1$, and its partition into $G^{+}_{1}$ and $G^{-}_{1}$ based on edge weights. Red-colored edges represent negative weights, while green-colored edges represent positive weights.}
    \label{fig:network-illustration}
\end{figure}

Hence, $G_i$ represents the local neighborhood of node $i$, while $G_i^+$ and $G_i^-$ separately capture its positive and negative connections, respectively. Figure~\ref{fig:network-illustration} illustrates an example of a weighted network and the corresponding star subgraph of a chosen node, showing its separation into positive and negative edge subsets.

To evaluate the relative strength of positive and negative connections associated with a node, we now define two measures.

The positive strength of node $i$, denoted by $S^+_{i}$, is defined as:
\begin{equation}\label{3eq:positive-strength}
    S^+_{i} = \frac{\sum_{\{i, j\} \in E(G^+_{i})} w_{ij}}{\sum_{\{i, j\} \in E(G_{i})} |w_{ij}|}.
\end{equation}

Similarly, the negative strength of node $i$, denoted by $S^-_{i}$, is defined as:
\begin{equation}\label{3eq:negative-strength}
    S^-_{i} = \frac{\sum_{\{i, j\} \in E(G^-_{i})} |w_{ij}|}{\sum_{\{i, j\} \in E(G_{i})} |w_{ij}|}.
\end{equation}

By construction, these two measures satisfy
\[
S^+_{i} + S^-_{i} = 1.
\]
A node is said to be positively dominated if $S^+_{i} \geq S^-_{i}$, and negatively dominated otherwise.

\subsubsection{SDG indicator classification model}

Sustainable development indicators are inherently interconnected, as progress in one area often influences outcomes in others, either positively or negatively. To represent and analyze these interdependencies systematically, we model the relationships among indicators as a network. In this network representation, each node corresponds to an individual indicator, and edges capture the statistical associations between pairs of indicators.

The strength of these associations is quantified using the Spearman rank correlation coefficient, which measures the degree of nonlinear monotonic relationship between two variables based on their ranked values. The coefficient, denoted by $\rho_{ij}$, ranges from $-1$ to $+1$. A positive correlation ($\rho_{ij} > 0$) indicates that as one indicator increases, the other tends to increase as well; values of $\rho_{ij}$ closer to $+1$ imply a stronger positive association between the two indicators. Such relationships are interpreted as synergies, where improvement in one indicator supports or reinforces progress in the other. Conversely, a negative correlation ($\rho_{ij} < 0$) suggests that as one indicator increases, the other tends to decrease; values closer to $-1$ indicate stronger opposition between their movements. This represents a trade-off, where progress in one dimension may come at the expense of another. When the correlation is equal to zero ($\rho_{ij} = 0$), it implies that the two indicators do not exhibit a monotonic relationship, meaning that the changes in one do not have an association with the changes in the other.

Before constructing the network, indicators with missing data or constant values over the analysis period (2000--2024) are excluded. 

For each country $k$, we construct a complete weighted network $G_{k} = (V_k, E_k, W_{k})$, where each node in $V_k$ represents a set consisting of SDG indicators of the country $k$, and each edge weight $w^{(k)}_{ij}$ corresponds to the Spearman rank correlation coefficient $\rho^{(k)}_{ij}$ between indicators $i$ and $j$ for that country. Thus,
$$
w^{(k)}_{ij} = \rho^{(k)}_{ij}, \quad \text{for all } \{i, j\} \in E_k.
$$

Using the notation introduced in Section~\ref{3 sec:network-measures}, we define the star subgraph $G_{ki}$ for each indicator $i$ in country $k$, consisting of node $i$ and all edges incident on it. This star subgraph can be further partitioned into two components:
$$
G_{ki}^{+} \text{ (synergistic subgraph)} \quad \text{and} \quad G_{ki}^{-} \text{ (trade-off subgraph)},
$$
corresponding respectively to positive and negative edge weights.

To quantify the relative importance of synergies and trade-offs associated with indicator $i$ in country $k$, we define the positive strength and negative strength as follows:
$$
S_{ki}^{+} = \frac{\sum_{\{i, j\} \in E(G_{ki}^{+})} \rho^{(k)}_{ij}}{\sum_{\{i, j\} \in E(G_{ki})} |\rho^{(k)}_{ij}|},
$$
$$
S_{ki}^{-} = \frac{\sum_{\{i, j\} \in E(G_{ki}^{-})} |\rho^{(k)}_{ij}|}{\sum_{\{i, j\} \in E(G_{ki})} |\rho^{(k)}_{ij}|}.
$$

Here, $S_{ki}^{+}$ represents the strength of synergies of indicator $i$ within country $k$, while $S_{ki}^{-}$ captures the strength of trade-offs. An indicator is said to be synergy-dominated if $S_{ki}^{+} \geq S_{ki}^{-}$, and trade-off dominated otherwise.

The derived measures are employed to classify SDG indicators into synergy-dominated and trade-off-dominated classes. To operationalize this classification, we apply statistical techniques to construct a probabilistic binary classification model. The objective of this study is to identify indicators that exhibit synergistic dominance, that is, indicators whose behavior has a positive impact on other indicators. For the dependent variable, we define $Y_{ki}$, as $$
Y_{ki} =
\begin{cases}
1, & \text{if } S_{ki}^+ \geq S_{ki}^- \\
0, & \text{otherwise}
\end{cases}
$$ for each country $1 \leq k \leq T$ and its indicators $1 \leq i \leq n_k$. Here, $T$ is the total number of countries considered in the study.  To effectively quantify the synergistic strength of each indicator, we use two network-theoretic metric-based predictors that separately capture direct and indirect effects. These predictors are defined as follows:

\begin{enumerate}
    \item To quantify the direct effect of a indicator node $i$ of the country $k$, we use a metric $X_{ki}^d$, which is the normalized number of positively weighted edges incident on that node; in particular, $X_{ki}^d$ is calculated by dividing the number of incident edges with positive weights by $n_k - 1$. This metric captures the immediate positive effect that node $i$ has on the SDG indicator interaction network. Normalization ensures that $X_{ki}^d$ lies between $0$ and $1$, allowing a meaningful comparison between countries with different numbers of indicators.

    \item To capture the indirect effect of a indicator node $i$, we employ the harmonic centrality measure $X_{ki}^h$, computed on the interaction network $G_k^{\text{strong}}$. The network $G_k^{\text{strong}}$ is an unweighted subgraph 
     of $G_k$, formed by retaining all vertices of $G_k$ but including only those edges whose weights are at least $0.8$. The harmonic centrality of a node $v_i \in V_k$ quantifies its accessibility from other nodes in the network and is defined as
\[
C_H(v_i) = \frac{1}{n_k-1} \sum_{v_i \ne v_j} \frac{1}{d(v_i, v_j)},
\]
where $n_k$ is the number of vertices in $G_k$, and $d(v_i, v_j)$ denotes the distance\footnote{The path between two nodes $v_i$ and $v_j$ is a sequence of distinct vertices $v_i=v_0,v_1,\dots,v_p=v_j$ such that each consecutive ordered pair $(v_{q-1}, v_q)$ is an edge in $E$. Here $p$ is the length of the path. The distance $d(v_i, v_j)$ between two nodes $v_i$ and $v_j$ is the length of the shortest such path. If there is no path from $v_i$ to $v_j$, then the distance between them is defined to be infinite.} between $v_i$ and $v_j$. If no path exists between the two nodes, $d(v_i, v_j)$ is considered infinite. Hence, a higher value of $X_{ki}^h$ indicates that node $i$ is more easily reachable from other nodes within the strong-interaction network $G_k^{\text{strong}}$.
 The threshold of $0.8$ is chosen to focus specifically on the flow of strong synergies through the network. By keeping only strong edges, we ensure that any indirect synergetic influence captured in the network arises from strong interactions, rather than being affected by weaker or uncertain synergies. Since the harmonic centrality of a node $i$ calculates the sum of reciprocals of shortest-path distances from every other node to node $i$, nodes that can reach $i$ in fewer steps of strong synergy contribute more to its score. This means that a node that is easily reachable from all other nodes through short sequences of strong connections will have a higher $X_{ki}^h$. 

\end{enumerate}




One of the simplest and robust classification methods is logistic regression, which belongs to the family of supervised machine learning approaches. We use the logistic regression model to estimate the conditional probability that an indicator $i$ for country $k$ exhibits synergistic behavior, using the direct effect $X_{ki}^d$ and the indirect effect $X_{ki}^h$ as predictors. The model is defined as
\[
\Pr(Y_{ki} = 1 \mid X_{ki}^d, X_{ki}^h) = \frac{1}{1 + \exp(-(\beta_0 + \beta_1 X_{ki}^d + \beta_2 X_{ki}^h))},
\]
where $\beta_0$, $\beta_1$, and $\beta_2$ are parameters to be estimated from the data.
 After fitting the model using maximum likelihood, we obtain the predicted probability:
$$
\hat{\Pr}(Y_{ki} = 1 \mid X_{ki}^d, X_{ki}^h) = \frac{1}{1 + \exp(-(\hat{\beta}_0 + \hat{\beta}_1 X_{ki}^d + \hat{\beta}_2 X_{ki}^h))},
$$
where $\hat{\beta}_0, \hat{\beta}_1 \text{ and } \hat{\beta}_2$ are the estimates of $\beta_0, \beta_1 \text{ and } \beta_2,$ respectively. The indicator is classified as synergy-dominated if the predicted probability is greater than or equal to 0.5, and trade-off-dominated otherwise.

\section{Results and Discussion}\label{3 section 3}

\subsection{Generic Model}
Based on the methodology described earlier, we constructed a complete weighted network $G_k$ for each country $k$ included in the study, after removing indicators with constant or missing values over the period 2000-2024. We then calculated the synergy and trade-off strengths for every node and categorized each indicator as either synergy-dominated or trade-off-dominated. Based on these findings, we used a logistic regression model to investigate the relationship between the likelihood of synergy dominance and network features, specifically the direct and indirect effects.

\subsubsection{Logistic Regression Model Output}

The logistic regression was performed using the combined data from all countries. Each country was first categorized based on its SDG Index score~\cite{sachs2025sdr}, which provides a measure of a country’s overall progress toward achieving the Sustainable Development Goals. Countries with SDG Index scores in the range of $80$ to $100$ were classified as ``best-performing,'' those with scores between $50$ and $80$ as ``moderate-performing,'' and those below $50$ as ``worst-performing.'' This categorization allows us to account for variation in development levels when building a predictive model. A detailed list of countries under each category is provided in Table~\ref{3 tab:sdg-categories} (See Appendix).

From each performance category, $80\%$ of the indicator data points were randomly selected to form the training dataset for the logistic regression. The remaining $20\%$ from each category were used as a testing dataset to evaluate the model’s predictive performance across different development contexts. This stratification ensures that the model is trained on a balanced representation of indicators across countries with varying SDG achievement levels.

The normalized direct effect ($X_{ki}^d$) and indirect effect ($X_{ki}^h$) were used as predictors, and the binary output variable $Y_{ki}$ indicated whether a indicator was synergy-dominated or trade-off-dominated. The Variance Inflation Factors (VIF) for both predictors were approximately 1.55, indicating that there is no serious multicollinearity. As noted in~\cite{james2013introduction}, a VIF value greater than 5 (or sometimes 10) indicates multicollinearity. 

 The estimated logistic regression model is:
\begin{equation}
\Pr(Y_{ki}= 1 \mid X_{ki}^d, X_{ki}^h) = \frac{1}{1 + \exp(-(-19.2031 + 39.0684 X_{ki}^d + 2.1742 X_{ki}^h))}.
\label{3 eq:logit_model}
\end{equation}

This model estimates the probability that a given indicator exhibits synergistic behavior, based on its direct and indirect influence within the interaction network. Indicators with predicted probability $\Pr(Y_{ki} = 1 \mid X_{ki}^d, X_{ki}^h) \geq 0.5$ are classified as synergy-dominated, whereas those with predicted probability less than $0.5$ are classified as trade-off-dominated.

When both the direct effect $X_{ki}^1$ and the indirect effect $X_{ki}^2$ are zero, the probability becomes:
$$
\Pr(Y_{ki} = 1 \mid X_{ki}^d = 0,X_{ki}^h = 0) = \frac{1}{1 + \exp(19.2031)} \approx 0.
$$
This implies that an indicator with neither direct nor indirect influence is extremely unlikely to be synergy-dominated. This is intuitive, as a node with no interaction or influence in the network is unlikely to contribute positively to broader progress.

Since both coefficients of $X_{ki}^d$ and $X_{ki}^h$ in the logistic model are positive, the probability of synergy-dominance increases with either higher direct effect or higher indirect effect. In particular, if two indicators have the same direct influence $(X_{ki}^d)$, then the one with a higher indirect influence $(X_{ki}^h)$ will have a higher predicted probability of being synergy-dominated. This demonstrates that indirect influence, measured through harmonic centrality in the strong interaction network, plays a non-redundant role. 


\subsection{Logistic Regression Model Evaluation}

Both predictors were statistically significant. The effect of $X_{ki}^d$ was very strong ($\hat{\beta} = 39.07$, SE = 1.47, 95\% CI [36.20, 41.94], $p < 0.001$). This suggests that it played a significant role in predicting the outcome. The small standard error and confidence interval of shorter length highlight a high degree of precision in the estimate. Also, the low $p$-value provides strong evidence opposing the null hypothesis of no effect. However, $X_{ki}^h$ showed a significant contribution, but with a smaller effect ($\hat{\beta} = 2.17$, SE = 0.93, 95\% CI [0.36, 3.99], $p = 0.019$). In this case, the larger standard error and confidence interval of longer length suggest more uncertainty around the effect size. Overall, $X_{ki}^h$ contributes positively since the interval excludes zero. A complete summary of the regression results is presented in Table~\ref{tab:logit_results}.

\begin{table}[h]
\centering
\caption{Logistic regression coefficients with standard errors, 95\% confidence intervals, and $p$-values.}
\label{tab:logit_results}
\begin{tabular}{lcccc}
\hline
\textbf{Predictor} & \textbf{$\hat{\beta}$} & \textbf{SE} & \textbf{95\% CI} & \textbf{$p$-value} \\
\hline
Intercept   & $-19.20$ & $0.71$ & $[-20.60,\;-17.81]$ & $<0.001$ \\
$X_{ki}^d$  & $39.07$  & $1.47$ & $[36.20,\;41.94]$   & $<0.001$ \\
$X_{ki}^h$  & $2.17$   & $0.93$ & $[0.36,\;3.99]$     & $0.019$ \\
\hline
\end{tabular}
\end{table}

To further evaluate predictive performance, we used the testing dataset composed of the remaining $20\%$ of data points from each SDG Index category. The predicted binary outputs were compared with the actual classifications of synergy-dominated $(Y_{ki} = 1)$ and trade-off-dominated $(Y_{ki} = 0)$ indicators.

The model achieved a high classification accuracy of $97.92\%$ on the testing data, indicating that it is highly effective in distinguishing between synergy- and trade-off-dominated indicators based on their direct and indirect network effects.


\begin{figure}[h]
    \centering
    \includegraphics[width=0.7\textwidth]{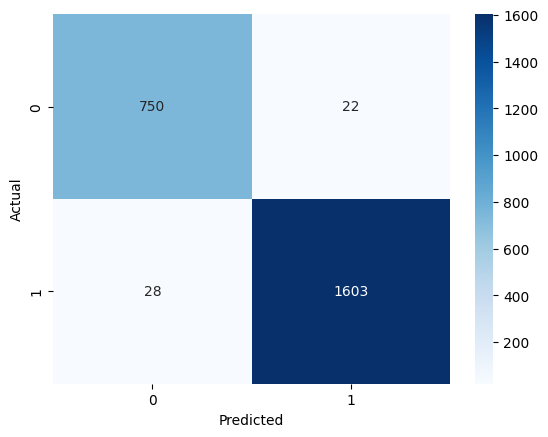}
    \caption{Confusion matrix}
    \label{3 fig2}
\end{figure}

According to the confusion matrix (Figure~\ref{3 fig2}), out of all synergy-dominated indicators, $1603$ were correctly predicted, and only $28$ were misclassified as trade-off-dominated.
 Among the trade-off-dominated indicators, $750$ were correctly predicted, while only $22$ were misclassified as synergy-dominated. The low number of misclassifications demonstrates the strong predictive performance and reliability of the proposed logistic regression model.

\subsection{Case Study: India}
\begin{figure}[h]
    \centering
    \includegraphics[width=1\textwidth]{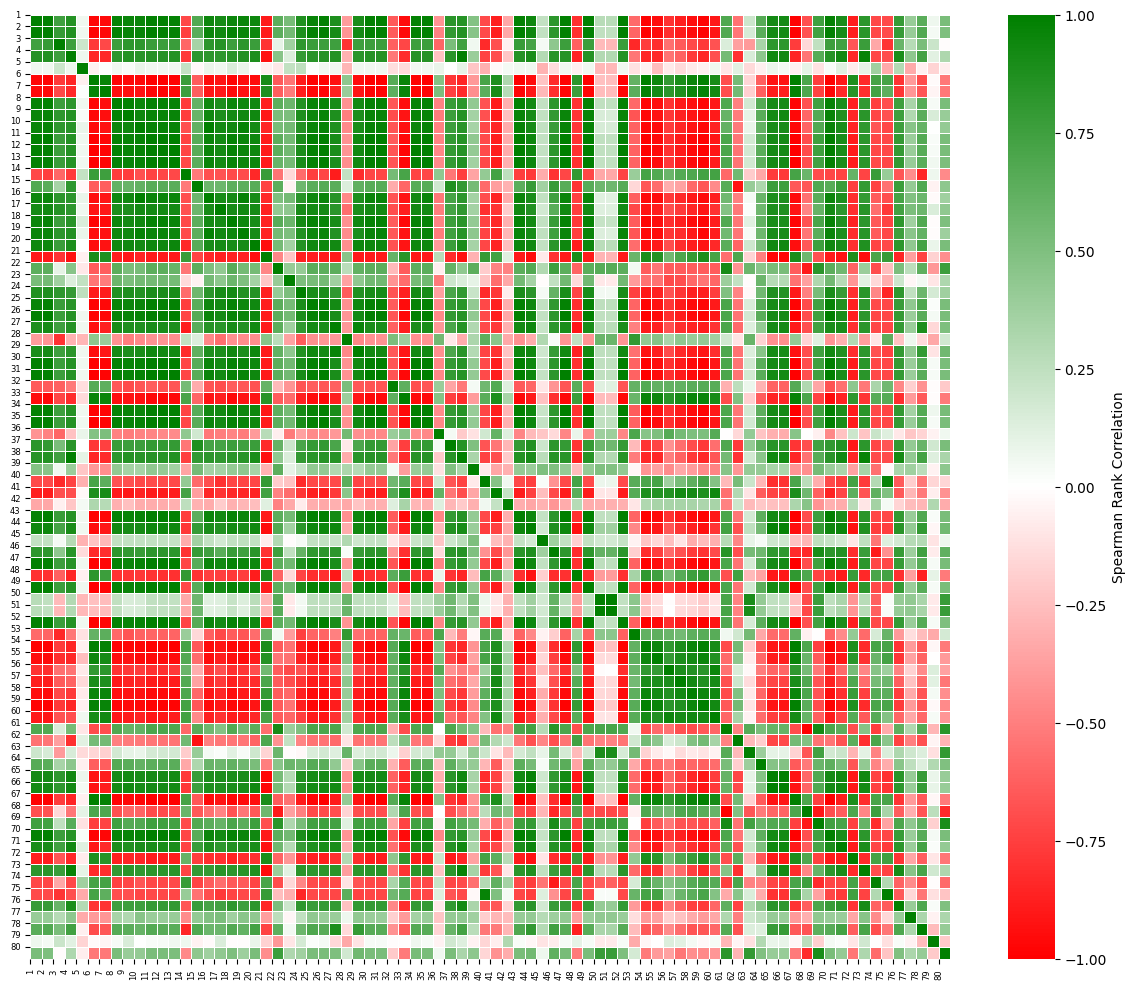}
    \caption{Spearman correlation heatmap of 80 sustainable development indicators for India, where dark red indicates strong negative correlation, dark green indicates strong positive correlation, and white indicates no correlation.}
    \label{3 correlation heat map}
\end{figure}

We conducted a case study on India to show the application of the suggested model. All of the nation's SDG indicators were included in the original dataset. To ensure meaningful Spearman correlations between indicator pairs, we removed indicators that remained constant throughout the period from 2000 to 2024, as well as those with missing values during this time. After the data cleaning process, 80 indicators were retained for further analysis. The complete list of these 80 SDG indicators considered in the study is presented in Table~\ref{3 tab:target_indicators} (See Appendix) and the Spearman correlations between targets are given in Figure~\ref{3 correlation heat map}.

To apply the logistic regression model to the Indian context, we first computed the normalized direct effect $(X_{ki}^d)$ and indirect effect $(X_{ki}^h)$ for each of the $80$ indicators, following the definitions given in the methodology. These values were then used to estimate the probability of an indicator being synergy-dominated using the logistic regression model given in Equation~\ref{3 eq:logit_model}. Indicators with $\Pr(Y_{ki} = 1 \mid X_{ki}^d, X_{ki}^h) \geq 0.5$ were considered as synergy dominated, while those with lower probabilities were considered as trade-off dominated. 

\begin{figure}[h]
    \centering
    \includegraphics[width=1\textwidth]{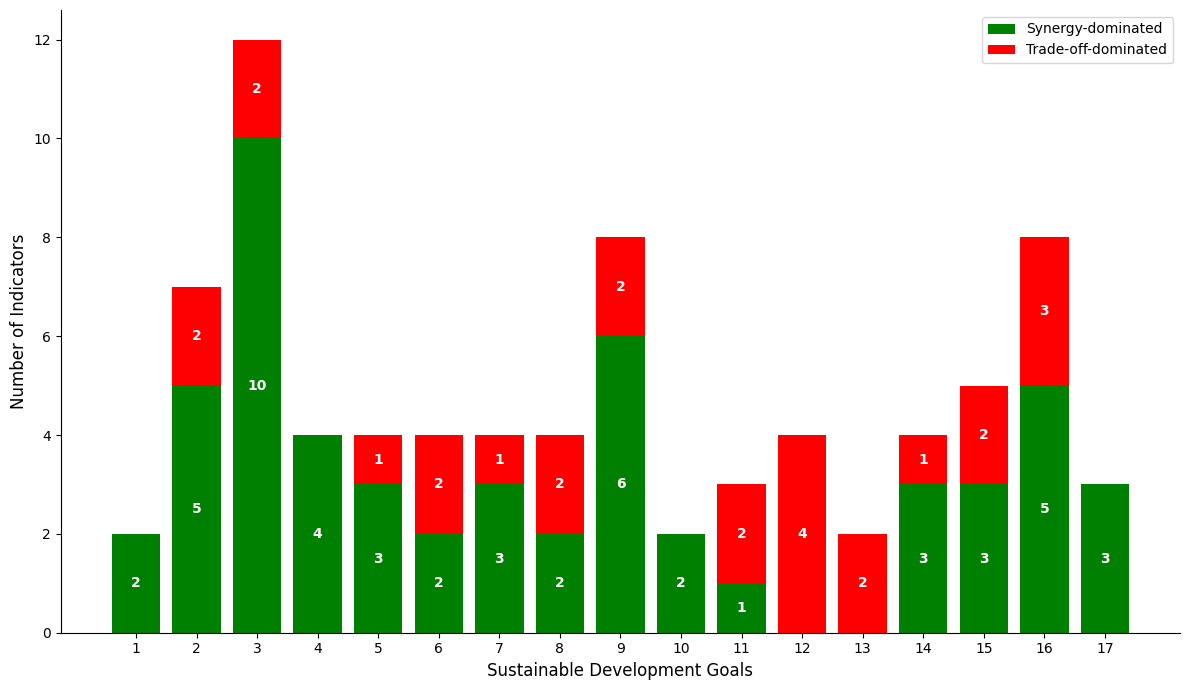}
    \caption{Grouped bar chart illustrating the number of synergy-dominated (green) and trade-off-dominated (red) indicators across each SDG for India.}
    \label{3 grouped bar chart}
\end{figure}

Of the $80$ indicators analyzed for India, $54$ were classified as synergy-dominated, and $26$ as trade-off-dominated. This suggests that the majority of indicators in the Indian context exhibit synergistic behavior, showcasing potential for integrated and mutually reinforcing policy interventions. However, the presence of a notable number of trade-off-dominated targets also cautions for careful policy consideration to avoid unintended negative consequences. To illustrate the distribution of synergy and trade-off-dominated indicators across the 17 Sustainable Development Goals, their counts are presented in Figure~\ref{3 grouped bar chart}.

In Figure~\ref{3 grouped bar chart}, we observe that SDG 3 (Good Health and Well-being) contains a large number of synergy-dominated indicators, meaning strong positive interlinkages with other goals and vast potential for progress in several areas simultaneously. However, SDG 12 (Responsible Consumption and Production) has the highest number of trade-off-dominated indicators, and none of them are synergy-dominated indicators. The five indicators of SDG 12 considered in the study are electronic waste (kg per capita), production-based air pollution, air pollution associated with imports, production-based nitrogen emissions, and nitrogen emissions associated with imports. 
This highlights the challenge in achieving SDG 12, where efforts to reduce pollution and manage waste can hinder economic or social objectives. 
For SDG 13 (Climate Action), both indicators $CO_2$ emissions from fossil fuel combustion and cement production, and GHG emissions embodied in imports, exhibit trade-off dominance, again revealing the conflict between economic and climate goals. SDG 11 (Sustainable Cities and Communities) also consists indicators that are trade-off dominated. One such indicator is access to improved water sources through piped connections. This appears counterintuitive, as expanding piped water access is generally considered a positive development outcome. However, in Figure~\ref{3 correlation heat map}, we observe that this indicator exhibits significant trade-offs with eleven indicators from SDG 3. Therefore, increasing urban water infrastructure may improve access, but it can also lead to unintended health issues. This observation is also supported by an empirical study~\cite{robert2023quality}, which shows that even when households have physical access to piped water, the quality of that water can be severely degraded due to pollution, poor waste management, or inadequate treatment infrastructure. 

Fishing by trawling or dredging is a synergy-dominated indicator for India. These fishing methods damage the seafloor by removing essential natural elements such as corals, shells, and sediments, which serve as shelter and breeding grounds for numerous fish. As a result, the fish population may decrease or fluctuate~\cite{national2002effects}. This implies that reducing such fishing techniques helps to maintain the ocean ecosystem. The ocean provides food, employment, and livelihoods for millions of people, particularly in coastal areas. Long-term economic growth, public health, and the preservation of fish stocks are supported by a healthy and sustainable ocean. It is also important to protect coastal communities from climate-related risks, such as rising sea levels, erosion, and extreme weather. Thus, preserving a healthy ocean ecosystem has numerous social and economic advantages~\cite{morales2024challenges,pendleton2020we}. Overall, these findings highlight the importance of understanding the interlinkages between targets to ensure that progress in one area does not regress the progress of another, but instead contributes to a more sustainable development trajectory.

\begin{figure}[h]
    \centering
    \includegraphics[width=0.9\textwidth]{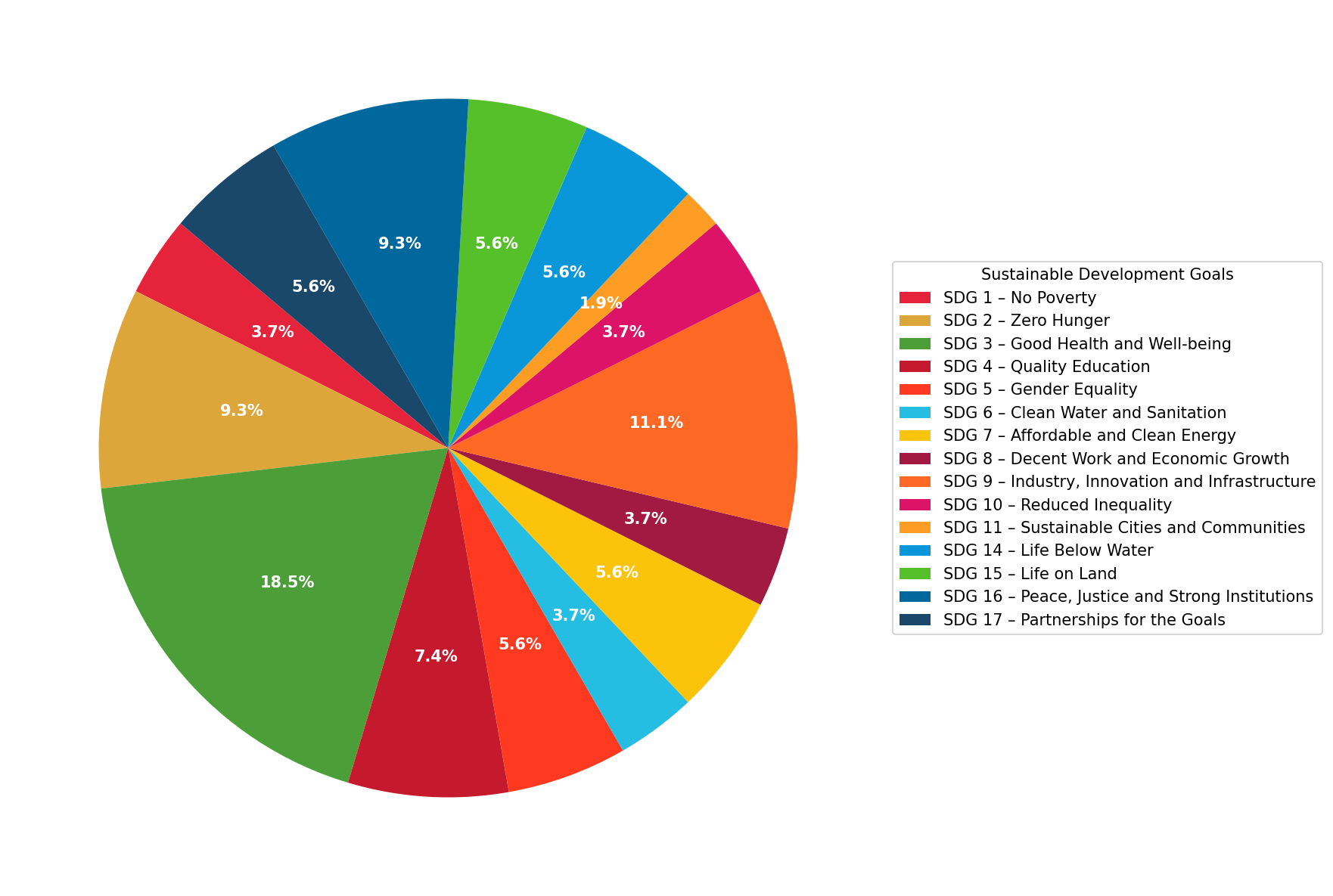}
    \caption{Pie chart showing the distribution of synergy-dominated indicators across the 17 Sustainable Development Goals for India.}
    \label{3 synergy pie}
\end{figure}

To better understand the distribution of synergy-dominated indicators in each goal, we present in Figure~\ref{3 synergy pie} the percentage share of these synergy-dominated indicators across the 17 SDGs. Unlike the grouped bar chart, which contrasts synergy and trade-off-dominated indicators, this pie chart focuses solely on the synergy-dominated indicators, thereby highlighting the relative contribution of each goal. 

\section{Conclusion}\label{3 section 4}
The Sustainable Development Goals are highly interconnected, and effective implementation requires a careful understanding of the interlinkages between the indicators. As policy implementations take place at the indicator level, identifying crucial indicators is more important than goals. 
We propose a generic network-based model to identify significant SDG indicators. The model estimates the likelihood that an indicator is synergy-dominated based on its direct and indirect synergistic effects by fitting a logistic regression model. We applied the model to the case of India to demonstrate its usefulness. The study revealed that SDG 3 (Good Health and Well-being) contains the highest number of synergy-dominated indicators, while SDG 12 (Responsible Consumption and Production) has the highest number of trade-off-dominated indicators and no synergy-dominated indicators, indicating difficulty in achieving SDG 12. Although we demonstrated for India, our model is general and can be applied to other countries for indicator classification.

While our model provides a general framework for identifying important SDG indicators, it has some limitations. First, it does not capture causal relationships between indicators, as it is completely dependent on observed correlations, which may capture associations but not directional influence. Second, the classification is based only on the strength of direct and indirect synergetic effects, without considering the feasibility or cost of implementing policies. Therefore, before acting on the synergy-dominated indicators, policymakers must consider other factors, such as resource availability and implementation constraints, even though the model identifies the influential targets.

\bibliographystyle{apalike}  
\bibliography{references}  
\clearpage
\appendix
\section*{Appendix}\label{Appendix}
\begin{table}[ht]
\centering
\renewcommand{\arraystretch}{1.2}
\caption{Categorization of countries based on SDG Index scores.}
\label{3 tab:sdg-categories}
\begin{tabular}{|p{15cm}|}
\hline
\multicolumn{1}{|c|}{\textbf{Worst-performing countries (SDG Index $<$ 50)}} \\
\hline
Afghanistan, Central African Republic, Chad, Democratic Republic of the Congo, Somalia, South Sudan, Sudan, Yemen\\
\hline
\multicolumn{1}{|c|}{\textbf{Moderate-performing countries (50 $\leq$ SDG Index $<$ 80)}} \\
\hline
Albania, Algeria, Angola, Argentina, Armenia, Australia, Azerbaijan, Bahamas, Bahrain, Bangladesh, Barbados, Belarus, Belize, Benin, Bhutan, Bolivia, Bosnia and Herzegovina, Botswana, Brazil, Brunei Darassalam, Bulgaria, Burkina Faso, Burundi, Cabo Verde, Cambodia, Cameroon, Canada, Chile, China, Colombia, Comoros, Costa Rica, C\^ote d'Ivoire, Cuba, Cyprus, Djibouti, Dominican Republic, Ecuador, Egypt, El Salvador, Eswatini, Ethiopia, Fiji, Gabon, Gambia, Georgia, Ghana, Greece, Guatemala, Guinea, Guinea-Bissau, Guyana, Haiti, Honduras, India, Indonesia, Iran, Iraq, Ireland, Israel, Jamaica, Jordan, Kazakhstan, Kenya, Korea Republic, Kuwait, Kyrgyz Republic, Lao PDR, Lebanon, Lesotho, Liberia, Lithuania, Luxembourg, Madagascar, Malawi, Malaysia, Maldives, Mali, Malta, Mauritania, Mauritius, Mexico, Moldova, Mongolia, Montenegro, Morocco, Mozambique, Myanmar, Namibia, Nepal, New Zealand, Nicaragua, Niger, Nigeria, North Macedonia, Oman, Pakistan, Panama, Papua New Guinea, Paraguay, Peru, Philippines, Qatar, Republic of the Congo, Romania, Russian Federation, Rwanda, S\~{a}o Tom\'{e} and Pr\'{\i}ncipe, Saudi Arabia, Senegal, Serbia, Sierra Leone, Singapore, South Africa, Sri Lanka, Suriname, Switzerland, Syrian Arab Republic, Tajikistan, Tanzania, Thailand, Togo, Trinidad and Tobago, Tunisia, T\"{u}rkiye, Turkmenistan, Uganda, Ukraine, United Arab Emirates, United States, Uruguay, Uzbekistan, Venezuela, Vietnam, Zambia, Zimbabwe\\
\hline
\multicolumn{1}{|c|}{\textbf{Best-performing countries (SDG Index $\geq$ 80)}} \\
\hline
Austria, Belgium, Croatia, Czechia, Denmark, Estonia, Finland, France, Germany, Hungary, Iceland, Italy, Japan, Latvia, Netherlands, Norway, Poland, Portugal, Slovak Republic, Slovenia, Spain, Sweden, United Kingdom\\
\hline
\end{tabular}
\end{table}

\begin{longtable}{clp{11cm}}
\caption{List of SDG Indicators}\label{3 tab:target_indicators} \\
\hline
No. & SDG & Indicator Description \\
\hline
\endfirsthead

\hline
No. & SDG & Indicator Description \\
\hline
\endhead

\hline
\endfoot

\hline
\endlastfoot

1 & SDG 1 & Poverty headcount ratio at \$2.15/day (2017 PPP, \%) \\
2 & SDG 1 & Poverty headcount ratio at \$3.65/day (2017 PPP, \%) \\
3 & SDG 2 & Prevalence of undernourishment (\%) \\
4 & SDG 2 & Prevalence of stunting in children under 5 years of age (\%) \\
5 & SDG 2 & Prevalence of wasting in children under 5 years of age (\%) \\
6 & SDG 2 & Prevalence of obesity, BMI $\geq 30$ (\% of adult population) \\
7 & SDG 2 & Human Trophic Level (best 2-3 worst) \\
8 & SDG 2 & Cereal yield (tonnes per hectare of harvested land) \\
9 & SDG 2 & Sustainable Nitrogen Management Index (best 0-1.41 worst)  \\
10 & SDG 3 & Maternal mortality rate (per 100,000 live births) \\
11 & SDG 3 & Neonatal mortality rate (per 1,000 live births) \\
12 & SDG 3 & Mortality rate, under-5 (per 1,000 live births) \\
13 & SDG 3 & Incidence of tuberculosis (per 100,000 population) \\
14 & SDG 3 & Age-standardized death rate due to cardiovascular disease, cancer, diabetes, or chronic respiratory disease in adults aged 30–70 years (\%) \\
15 & SDG 3 & Traffic deaths (per 100,000 population) \\
16 & SDG 3 & Life expectancy at birth (years) \\
17 & SDG 3 & Adolescent fertility rate (births per 1,000 females aged 15 to 19)  \\
18 & SDG 3 & Births attended by skilled health personnel (\%) \\
19 & SDG 3 & Surviving infants who received 2 WHO-recommended vaccines (\%) \\
20 & SDG 3 & Universal health coverage (UHC) index of service coverage (worst 0-100 best) \\
21 & SDG 3 & Subjective well-being (average ladder score, worst 0-10 best) \\
22 & SDG 4 & Participation rate in pre-primary organized learning (\% of children aged 4 to 6) \\
23 & SDG 4 & Net primary enrollment rate (\%) \\
24 & SDG 4 & Lower secondary completion rate (\%) \\
25 & SDG 4 & Literacy rate (\% of population aged 15 to 24) \\
26 & SDG 5 & Demand for family planning satisfied by modern methods (\% of females aged 15 to 49) \\
27 & SDG 5 & Ratio of female-to-male mean years of education received (\%) \\
28 & SDG 5 & Ratio of female-to-male labor force participation rate (\%) \\
29 & SDG 5 & Seats held by women in national parliament (\%) \\
30 & SDG 6 & Population using at least basic drinking water services (\%) \\
31 & SDG 6 & Population using at least basic sanitation services (\%) \\
32 & SDG 6 & Freshwater withdrawal (\% of available freshwater resources) \\
33 & SDG 6 & Scarce water consumption embodied in imports ($m^3H_2O$ eq/capita) \\
34 & SDG 7 & Population with access to electricity (\%) \\
35 & SDG 7 & Population with access to clean fuels and technology for cooking (\%) \\
36 & SDG 7 & $CO_2$ emissions from fuel combustion per total electricity output $(MtCO_2/TWh)$  \\
37 & SDG 7 & Renewable energy share in total final energy consumption (\%) \\
38 & SDG 8 & Adults with an account at a bank or other financial institution or with a mobile-money-service provider (\% of population aged 15 or over) \\
39 & SDG 8 & Unemployment rate (\% of total labor force, ages 15+) \\
40 & SDG 8 & Fundamental labor rights are effectively guaranteed (worst 0–1 best) \\
41 & SDG 8 & Fatal work-related accidents embodied in imports (per million population) \\
42 & SDG 9 & Rural population with access to all-season roads (\%) \\
43 & SDG 9 & Population using the internet (\%) \\
44 & SDG 9 & Mobile broadband subscriptions (per 100 population) \\
45 & SDG 9 & Logistics Performance Index: Quality of trade and transport-related infrastructure (worst 1-5 best) \\
46 & SDG 9 & The Times Higher Education Universities Ranking: Average score of top 3 universities (worst 0-100 best) \\
47 & SDG 9 & Articles published in academic journals (per 1,000 population) \\
48 & SDG 9 & Expenditure on research and development (\% of GDP) \\
49 & SDG 9 & Total patent applications by applicant's origin (per million population) \\
50 & SDG 10 & Gini coefficient \\
51 & SDG 11 & Proportion of urban population living in slums (\%) \\
52 & SDG 11 & Annual mean concentration of particulate matter of less than 2.5 microns in diameter (PM2.5) ($\mu g / m^3$) \\
53 & SDG 11 & Access to improved water source, piped (\% of urban population) \\
54 & SDG 12 & Electronic waste (kg/capita) \\
55 & SDG 12 & Production-based air pollution (DALYs per 1,000 population) \\
56 & SDG 12 & Air pollution associated with imports (DALYs per 1,000 population) \\
57 & SDG 12 & Production-based nitrogen emissions (kg/capita) \\
58 & SDG 12 & Nitrogen emissions associated with imports (kg/capita) \\
59 & SDG 13 & $CO_2$ emissions from fossil fuel combustion and cement production (tCO2/capita) \\
60 & SDG 13 & GHG emissions embodied in imports ($tCO_2$/capita) \\
61 & SDG 14 & Ocean Health Index: Clean Waters score (worst 0-100 best) \\
62 & SDG 14 & Fish caught from overexploited or collapsed stocks (\% of total catch) \\
63 & SDG 14 & Fish caught by trawling or dredging (\%) \\
64 & SDG 14 & Fish caught that are then discarded (\%) \\
65 & SDG 15 & Mean area that is protected in terrestrial sites important to biodiversity (\%) \\
66 & SDG 15 & Mean area that is protected in freshwater sites important to biodiversity (\%) \\
67 & SDG 15 & Red List Index of species survival (worst 0-1 best)  \\
68 & SDG 15 & Permanent deforestation (\% of forest area, 3-year average) \\
69 & SDG 15 & Imported deforestation ($m^2$/capita) \\
70 & SDG 16 & Homicides (per 100,000 population) \\
71 & SDG 16 & Crime is effectively controlled \\
72 & SDG 16 & Unsentenced detainees (\% of prison population) \\
73 & SDG 16 & Corruption Perceptions Index (worst 0-100 best) \\
74 & SDG 16 & Press Freedom Index (worst 0-100 best) \\
75 & SDG 16 & Access to and affordability of justice (worst 0–1 best) \\
76 & SDG 16 & Timeliness of administrative proceedings (worst 0 - 1 best) \\
77 & SDG 16 & Expropriations are lawful and adequately compensated (worst 0 - 1 best) \\
78 & SDG 17 & Government spending on health and education (\% of GDP) \\
79 & SDG 17 & Other countries: Government revenue excluding grants (\% of GDP) \\
80 & SDG 17 & Statistical Performance Index (worst 0-100 best) \\

\end{longtable}

\end{document}